\documentclass[article,nojss]{jss}

\usepackage{orcidlink,thumbpdf,lmodern}
\usepackage{amssymb,amsmath,amsthm,mathtools,bm}
\usepackage{upquote}

\usepackage{microtype}
\UseMicrotypeSet[protrusion]{basicmath} 

\usepackage{booktabs}
\usepackage{caption}
\usepackage{subcaption}
\usepackage{siunitx}

\usepackage[linesnumbered,vlined,ruled]{algorithm2e}

\usepackage{enumitem}

\makeatletter
\let\natwidth\Gin@nat@width
\makeatother

\DeclareMathOperator*{\minimize}{minimize}
\DeclareMathOperator{\sign}{sign}

\newcommand{\fct}[1]{\code{#1()}}
\newcommand{\dataset}[1]{\code{#1}}

\newcommand{\link}{g}
\newcommand{\ilink}{g^{-1}}

\newcommand{\cC}{\mathcal{C}}

\newcommand{\myurl}[1]{\href{https://#1}{\nolinkurl{#1}}}

\author{
  Johan Larsson~\orcidlink{0000-0002-4029-5945}\\
  University of Copenhagen \And Ma\l{}gorzata
  Bogdan~\orcidlink{0000-0002-0657-4342}\\
  University of Wroc\l{}aw \And Krystyna
  Grzesiak~\orcidlink{0000-0003-2581-7722}\\
  University of Wroc\l{}aw \AND Mathurin
  Massias~\orcidlink{0000-0002-8950-0356}\\
  Inria, ENS de Lyon; CNRS, Lyon \And Jonas
  Wallin~\orcidlink{0000-0003-0381-6593}\\
  Lund University
}
\Plainauthor{Johan Larsson, Malgorzata Bogdan, Krystyna Grzesiak, Mathurin
  Massias, Jonas Wallin}

\title{
  Efficient Solvers for SLOPE in \proglang{R}, \proglang{Python},
  \proglang{Julia}, and \proglang{C++}
}
\Plaintitle{Efficient Solvers for SLOPE in R, Python, Julia, and C++}
\Shorttitle{Efficient Solvers for SLOPE}

\Abstract{
  \noindent We present a suite of packages in \proglang{R}, \proglang{Python},
  \proglang{Julia}, and \proglang{C++} that efficiently solve the Sorted L-One
  Penalized Estimation (SLOPE) problem. The packages feature a highly efficient
  hybrid coordinate descent algorithm that fits generalized linear models (GLMs)
  and supports a variety of loss functions, including Gaussian, binomial,
  Poisson, and multinomial logistic regression. Our implementation is designed
  to be fast, memory-efficient, and flexible. The packages support a variety of
  data structures (dense, sparse, and out-of-memory matrices) and are designed
  to efficiently fit the full SLOPE path as well as handle cross-validation of
  SLOPE models, including the relaxed SLOPE. We present examples of how to use
  the packages and benchmarks that demonstrate the performance of the packages
  on both real and simulated data and show that our packages generally
  outperform existing implementations of SLOPE in terms of speed.
}

\Plainkeywords{SLOPE, OWL, regularization, generalized linear models, R, Python,
  Julia, C++}

\Address{
  Johan Larsson\\ Department of Mathematical Sciences\\ Faculty of Science\\
  University of Copenhagen\\ Universitetsparken 5\\ 2100 København Ø, Denmark\\
  E-mail: \email{johan@jolars.co}\\ URL: \url{https://jolars.co/}
}

\begin{document}

\section{Introduction}

Sorted L-One Penalized Estimation (SLOPE)~\citep{bogdan2013, zeng2014,
bogdan2015} is a type of regularized regression that consists of solving the
following convex optimization problem:
\begin{equation}
  \label{eq:slope}
  \minimize_{\beta_0 \in \mathbb{R},\beta \in \mathbb{R}^p} \Big( P(\beta_0,\beta) = F(\beta_0, \beta) + \alpha J_\lambda(\beta) \Big)
\end{equation}
where \(P\) is the primal problem, \(F\) is the loss function, \(\alpha\) is a
parameter that controls the strength of regularization, and \(\lambda\) is a
non-increasing sequence of penalty weights. \(J\) is the \emph{sorted $\ell_1$
norm}, defined as
\begin{equation}
  J_\lambda(\beta) = \sum_{j=1}^p \lambda_j |\beta_{(j)}|, \quad \text{where}\quad |\beta_{(1)}| \geq |\beta_{(2)}| \geq \ldots \geq |\beta_{(p)}|.
\end{equation}
We will assume that \(F\) takes the following form:
\[
  F( \beta_0, \beta ) = \frac{1}{n} \sum_{i=1}^n f( \beta_0 + x_i^\intercal \beta , y_i),
\]
where \(f\) is a smooth, convex function and \(x_i\) is the \(i\)th row of the
\(n \times p\) design matrix \(X\). Throughout the paper, we use scalar-response
notation and denote a row of \(X\) by \(x_i\) and a column by \(x_j\). For
multinomial regression with \(m\) classes, \(y_i\) and \(\beta_0\) instead have
\(m - 1\) components, \(\beta\) is a \(p \times (m - 1)\) matrix, and the sorted
\(\ell_1\) norm acts on its vectorization. We let
\((\hat{\beta}_0, \hat{\beta})\) denote a solution to the problem in
\autoref{eq:slope}.

SLOPE is a generalization of both the lasso\footnote{The lasso is attained by
taking a constant \(\lambda\).}~\citep{santosa1986, donoho1994, donoho1995,
tibshirani1996} and OSCAR\footnote{OSCAR is attained by setting \(\lambda\) to
be a linear sequence, where \(\lambda_j = \theta_1 + \theta_2(p - j)\) with
\(\theta_1, \theta_2 \geq 0\)~\citep{figueiredo2014}.} (octagonal shrinkage and
clustering algorithm for regression)~\citep{bondell2008}. One of the most
important properties of SLOPE is that it can cluster coefficients by setting
them to the same magnitude~\citep{figueiredo2016, hejny2025, bogdan2026}. This
is a natural consequence of the sorted \(\ell_1\) norm, stemming from the fact
that the contribution to the norm of a given coefficient increases
disproportionately if it changes order. This also allows SLOPE to better recover
the ordering pattern in the solution.

Like the lasso, SLOPE is a convex but non-smooth optimization problem. And since
the pool-adjacent-violators algorithm~(PAVA)~\citep{barlow1972} can be used to
efficiently\footnote{At an average \(p \log p\) rate, due to the limiting
sorting operation.} compute the proximal operator of the sorted \(\ell_1\) norm,
it is possible to use a wide range of proximal algorithms, such as proximal
gradient descent, to solve SLOPE\@. This also includes accelerated methods such
as FISTA~\citep{beck2009}, which was for instance used by \citet{bogdan2015}.
Other possibilities include proximal Newton~\citep{lee2014} and the alternating
direction method of multipliers (ADMM) method~\citep{boyd2010}.

For similar problems such as the lasso and elastic net\citep{zou2005}, however,
these aforementioned methods have generally found themselves outperformed by
coordinate descent methods~\citep{friedman2007, friedman2010}, which optimize
one coefficient at a time. Several efficient implementations of coordinate
descent algorithms exists for the lasso, elastic net, and other \(\ell_1\)-based
regularization methods. For \proglang{R}, there is for instance
\pkg{glmnet}~\citep{friedman2010}, which fits generalized linear models,
including Cox-proportional hazards models, with elastic net regularization.
Other notable packages are \pkg{ncvreg}~\citep{breheny2011}, which in addition
to the elastic net also supports non-convex penalties such as
SCAD~\citep{fan2001} and MCP~\citep{zhang2010}. Another interesting package is
\pkg{biglasso}~\citep{zeng2021}, which supports out-of-memory data to fit lasso
models to large datasets. Finally, there is also \pkg{lars}~\citep{efron2004},
which fits exact lasso solution paths using the least angle regression (LARS)
algorithm. For \proglang{Python}, there is support for the elastic net in the
widely used \pkg{scikit-learn}~\citep{pedregosa2011} module, which beyond the
coordinate descent algorithm also features LARS and alternative solvers for the
lasso and elastic net. For \proglang{Julia}, there is generally less support for
regularized regression. There exists a wrapper for \pkg{glmnet} in
\pkg{GLMNet.jl}~\citep{kornblith2024a} and a native \proglang{Julia} package in
\pkg{Lasso.jl}~\citep{kornblith2024}.

Unfortunately, coordinate descent requires that the objective is separable in
\((\beta_0, \beta)\), which is not the case in SLOPE due to the permutations
involved in the sorted \(\ell_1\) norm, which means that coordinate descent
cannot be used directly. This problem was, however, overcome by
\citet{larsson2023}, who invented a hybrid combination of proximal gradient and
coordinate descent. The algorithm alternates between proximal gradient descent
steps on the full problem and coordinate descent on a collapsed problem
corresponding to the current cluster structure, to achieve robust and fast
convergence.

In this paper we make this algorithm available to a wide audience by presenting
a collection of packages in \proglang{R}~\citep{rcoreteam2025},
\proglang{Python}~\citep{pythonsoftwarefoundation2025},
\proglang{Julia}~\citep{bezanson2017}, and \proglang{C++}. For \proglang{Julia}
and \proglang{Python}, these are, to the best of our knowledge, the first and
only available implementations of SLOPE\@. For \proglang{R}, our package was the
first implementation of SLOPE, although there is now also support for SLOPE
through the \pkg{sgs}~\citep{feser2023} and \pkg{grpSLOPE}~\citep{brzyski2018}
packages. Additionally, an implementation for \proglang{MATLAB} is also
available from \citet{candes2013} and is also included in
TFOCS~\citep{becker2011}.

\subsection{Outline of the paper}

In \autoref{sec:math-details}, we introduce the statistical problem that our
packages solve, namely generalized linear models (GLMs) regularized with the
sorted \(\ell_1\) norm (SLOPE)\footnote{SLOPE is sometimes used only as the
procedure that uses quadratic loss, but here we adopt a more general terminology
and let SLOPE be defined for any loss function.}, and provide a brief overview
of the mathematical properties of GLMs and the particular properties of SLOPE.
We discuss the optimization problem and describe the hybrid coordinate descent
algorithm we use for solving SLOPE\@. In \autoref{sec:implementation-details},
we provide a detailed overview of the software implementations, highlighting
technical aspects such as memory management, parallelization, and convergence
criteria. In \autoref{sec:examples}, we showcase how our packages work in
practice, providing examples of fitting models, plotting, and performing
cross-validation. Finally, in \autoref{sec:discussion}, we summarize the
contributions of this paper and discuss future work.

\section{Mathematical details}
\label{sec:math-details}

In this section, we provide a brief overview of the mathematical details of the
SLOPE optimization problem, including the objective and the hybrid coordinate
descent algorithm used to solve it. We also discuss the convergence criteria
used to determine when the algorithm has converged, and the path fitting
procedure used to compute the full regularization path for SLOPE.

\subsection{Generalized linear models}

Our packages are designed to solve SLOPE for generalized linear models (GLMs),
in which the response \(y_i\) is modelled as a random variable from an
exponential family, and is assumed to depend conditionally on the linear
predictor \(\eta_i = x_i^\intercal \beta + \beta_0\) via
\(\E(y_i \mid \eta_i) = \ilink(\eta_i)\), where \(\ilink\) is the inverse link
function. This gives rise to the following form of the loss function:
\[
  F( \beta_0, \beta ) = \frac{1}{n} \sum_{i=1}^n f( \eta_i, y_i),
\]
where \(f\) is the contribution of the \(i\)th observation to the negative
log-likelihood of the model. To estimate the parameters of a generalized linear
model, \((\beta_0, \beta)\), we minimize the loss function \(F\) with respect to
\((\beta_0, \beta)\).

A special property of \(f\) is that its partial derivative with respect to
\(\eta\) is
\[
  \frac{\partial}{\partial \eta_i} f( \eta_i, y_i) = \ilink ( \eta_i) - y_i
                                                   = r_i,
\]
where we define \(r_i\) to be the \emph{generalized residual}. As a consequence,
the gradient of the loss function with respect to \(\beta\) can be expressed as
\[
  \frac{\partial}{\partial \beta_j} F( \beta_0, \beta )
  = \frac{1}{n} \sum_{i=1}^n x_{ij} \frac{\partial}{\partial \eta_i} f( \eta_i, y_i)
  = \frac{1}{n} \sum_{i=1}^n x_{ij} r_i.
\]

We summarize the loss, link, and inverse link functions for the GLMs supported
by the SLOPE packages in \autoref{tab:glm}.

\begin{table}[t!]
  \centering
  \begin{tabular}{lccc}
    \toprule
    Model       &                                    \(f(\eta, y)\)                                     &                     \(\link(\mu)\)                      &                                 \(\ilink(\eta)\)                                  \\
    \midrule
    Gaussian    &                              \(\frac{1}{2}(y - \eta)^2\)                              &                         \(\mu\)                         &                                     \(\eta\)                                      \\
    \addlinespace
    Binomial    &                             \(\log(1 + e^\eta) - \eta y\)                             &       \(\log \left( \frac{\mu}{1 - \mu} \right)\)       &                           \(\frac{e^\eta}{1 + e^\eta}\)                           \\
    \addlinespace
    Poisson     &                                  \(e^\eta - \eta y\)                                  &                      \(\log(\mu)\)                      &                                    \(e^\eta\)                                     \\
    \addlinespace
    Multinomial & \(\log \left( 1 + \sum_{j=1}^{m-1} e^{\eta_j} \right) - \sum_{k=1}^{m-1} y_k \eta_k\) & \(\left( \log \frac{\mu_k}{\mu_m} \right)_{k=1}^{m-1}\) & \(\left( \frac{e^{\eta_k}}{1 + \sum_{j=1}^{m-1} e^{\eta_j}} \right)_{k=1}^{m-1}\) \\
    \bottomrule
  \end{tabular}
  \caption{
    Loss functions, link functions, and inverse link functions for generalized
    linear models in the SLOPE package. Note that in the case of multinomial
    logistic regression, the input is vector-valued and
    \(\mu_m = 1 - \sum_{k=1}^{m-1} \mu_k\) is the mean of the reference class.
  }
  \label{tab:glm}
\end{table}

A particular case of interest is the multinomial logistic regression model. Many
implementations of regularized multinomial logistic regression models, such as
those by \citet{friedman2010} and \citet{fercoq2015}, use the \emph{redundant}
\(m\)-class formulation. Here, however, we have opted to use the non-redundant
formulation of the loss function, with the last class serving as the reference
category. This choice slightly complicates the notation and leads to a more
complex formulation for the dual problem. For SLOPE, however, this is a more
natural choice, since it generalizes to the binary case as well (in which the
last class is also implicit).\footnote{The non-redundant formulation requires a
\(\lambda\) sequence of length \((m - 1)p\). It would therefore not reduce to
the binary formulation if we instead used \(m\) coefficient vectors.}
Furthermore, it also relieves the problem of the parameter ambiguity that
affects the redundant formulation, where rows of the coefficient matrix can be
shifted without changing the value of \(F(\beta_0, \beta)\). This also means
that the redundant formulation needs to be accompanied by a bounds adjustment
step~\citep{friedman2010}. This is trivial in the case of the lasso and ridge,
and only slightly more complicated for the elastic net. For SLOPE, however, the
situation is much more involved since rows of the coefficient matrix cannot be
arbitrarily shifted, since this could affect the entire clustering structure.

\subsection{Hybrid algorithm}

The primary algorithm of the \pkg{SLOPE} packages is the hybrid coordinate
descent algorithm by \citet{larsson2023}. Since it is described in detail there,
we will only summarize its key points here and highlight some of the
improvements that we have made to the original algorithm. The basic idea is to
perform the coordinate descent updates on the clusters' coefficients rather than
directly on the coefficients.

For a fixed $\beta$ such that $|\beta_j|$ takes $m$ distinct values, we let
\(\mathcal{C}_1, \mathcal{C}_2 , \dots, \mathcal{C}_m\) and
\(c_1, c_2, \dots, c_m\) be the indices and coefficients, respectively, of the
\(m\) clusters of $\beta$, such that $\mathcal{C}_i = \{j : |\beta_j| = c_i\}$
and $c_1 > c_2 > \cdots > c_m \geq 0.$

In our algorithm, we fix the current cluster, e.g. $\mathcal{C}_k$, and update
all coefficients belonging to it in a single step. By itself, this algorithm is
not guaranteed to converge, since it can only reorder or merge clusters. To
circumvent this, the algorithm must be combined with proximal gradient descent
steps. These steps are able to split the clusters, which eventually means that
the full algorithm converges to the correct clusters and global minimum.

We have outlined the algorithm in \autoref{alg:hybrid}. Unlike the
implementation in \citet{larsson2023}, this algorithm is extended to any
generalized linear model---not just the Gaussian case. To do so, we use an
iteratively reweighted least-squares (IRLS) approach, determining weights \(w\)
and a working response \(z\) after each proximal gradient descent step. We then
run the coordinate descent algorithm for a fixed number of iterations.
Convergence is monitored using a duality-based stopping
criterion~(\autoref{sec:convergence-criteria-details}), which we check after
each gradient computation. To avoid complicating the presentation, we have
omitted some details of the algorithm here.

\begin{algorithm}[tp]
  \caption{
    The Hybrid coordinate descent algorithm for generalized linear models, using
    the IRLS where weights \(w\) and a working response \(z\) are computed after
    a proximal gradient descent step. We describe the cyclical version of the
    algorithm here and have omitted the intercept for simplicity.
    \(c^{\setminus k}\) is a version of the vector \(c\) with the \(k\)th
    coefficient omitted and \(T\) is the SLOPE thresholding operator, which we
    have illustrated in \autoref{fig:slope-thresholding}.
  }
  \label{alg:hybrid}
  \SetKwInOut{Input}{input}
  \SetKwComment{tcc}{$\triangleright$ }{}
  \SetCommentSty{textit}

  \Input{%
    \(X \in \mathbb{R}^{n \times p}\),
    \(y \in \mathbb{R}^n\),
    \(\lambda \in \{\mathbb{R}^p : \lambda_1 \geq \lambda_2 \geq \cdots > 0\}\),
    \(v \in \mathbb{N}\)
  }

  \Repeat{\(P(\beta_0, \beta) - D(\theta) \leq \varepsilon|P(\beta_0, \beta)|\)}{
    set \(t\) with backtracking line search\;

    \(\beta \gets \beta - t \nabla F(\beta_0, \beta)\)\;

    \For{\(i \gets 1,\dots,n\)}{
      \(\eta_i \gets x_i^T \beta + \beta_0\)\;
      \(w_i \gets \frac{\partial^2}{\partial \eta_i^2} f(\eta_i, y_i)\)\tcc*{Weights
        for IRLS}
      \(z_i \gets \eta_i - \frac{r_i(\eta_i, y_i)}{w_i}\)\tcc*{Working response
        for IRLS}
    }
    update \(c\), \(\mathcal{C}\)\;
    \For{\texttt{it} \(\gets 1,\dots,\)\texttt{cd\_maxit}}{
      \(k \gets 1\)\;
      \(\beta^\text{old} \gets \beta\)\;

      \While{\(k \leq \lvert \mathcal{C} \rvert\)}{
        \(\tilde x \gets \sum_{j \in \cC_k} x_j \sign \beta_j\)\;
        \(\tilde{r} \gets X \beta + \beta_0 - z\)\tcc*{Residuals}
        \(\gamma \gets \frac{1}{n} \sum_{i=1}^n w_i \tilde{x}_i \tilde{r}_i\)\tcc*{Gradient}
        \(\xi \gets \frac{1}{n} \sum_{i=1}^n w_i \tilde{x}_i^2\)\tcc*{Hessian}
        \(\tilde {c} \gets T(c_k \xi - \gamma; \xi, c^{\setminus k}, \lambda)\)\;
        \(\beta_{\cC_k} \gets \tilde{c} \sign(\beta_{\cC_k})\)\; Update \(c\),
        \(\mathcal{C}\)\; \(k \gets k + 1\)\;
      }

      \If{\(P(\beta) \geq P(\beta^\text{old})\)}{
        \(\beta \gets \beta^\text{old}\)\; break\;
      }
    }
  }

\end{algorithm}

\begin{figure}[t]
  \centering
  \includegraphics[width=\natwidth]{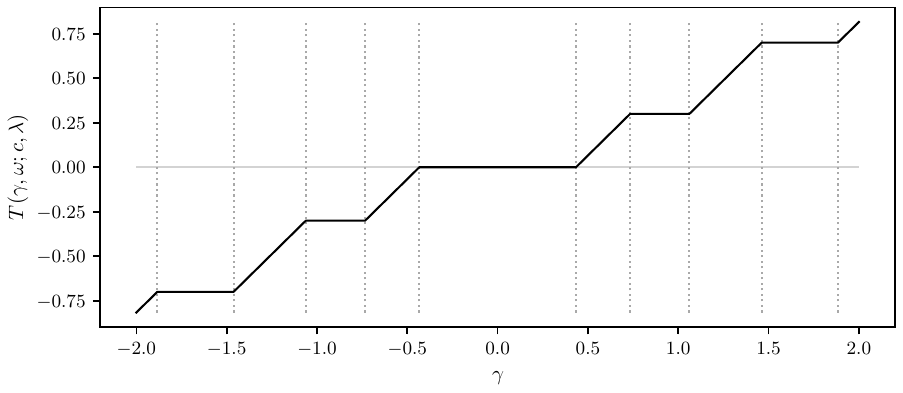}
  \caption{
    An illustration of the SLOPE thresholding operator for
    \(\beta = [0.5, - 0.5, 0.3, 0.7]^\intercal\), and thus
    $c = \{0.7, 0.5, 0.3\}$, where we consider an update for the second cluster,
    \(C_2 = \{1, 2\}\). We have omitted the particular value of $\omega$ for
    brevity. The example is adapted from \citet{larsson2023}. Across regions
    where the function is constant, the operator sets the result to be either
    exactly 0 or merges the cluster with another cluster, setting their
    magnitudes to be equal.
  }
  \label{fig:slope-thresholding}
\end{figure}

The hybrid nature of the algorithm is not only useful for enabling the
coordinate descent steps, but also comes with benefits due to the inclusion of
the PGD steps. Proximal gradient descent algorithms are known to converge under
more general assumptions than coordinate descent algorithms~\citep{wright2015},
which means that the PGD part of the algorithm can serve as a fallback in case
the coordinate descent part does not converge. In practice, we store the current
solution before updating and revert to the previous solution if the coordinate
descent step does not improve the primal objective.

The implementation in \citet{larsson2023} used a type of cyclical coordinate
descent in which the algorithm iterated over the clusters in descending order by
their coefficients' magnitudes. Although simple to implement and efficient for
many problems, cyclical coordinate descent is known to suffer from convergence
issues in some cases~\citep{wright2015}, which we found to be the case for SLOPE
as well. For the current implementation, we therefore include, and default to, a
version using random permutations, which is slightly slower but more robust.

\subsection{Convergence criteria}

Our packages use a duality-based stopping criterion, providing an upper bound on
suboptimality at convergence. We transform the primal problem,
\(P(\beta_0, \beta)\), into a constrained formulation and derive a dual problem,
\(D(\delta)\), that allows us to compute a duality gap. As a stopping criterion,
we use the relative duality gap:
\[
  P( \beta_0, \beta ) - D( \delta ) \leq \varepsilon | P( \beta_0, \beta ) |,
\]
where \(\varepsilon > 0\) is the user-defined tolerance level. This provides a
reliable, solver-independent measure of convergence. The complete derivation of
the dual problem and the calculation of the duality gap is provided in
\autoref{sec:convergence-criteria-details}.

\subsection{Path fitting}

Since optimal settings of \(\alpha\) are only available under strict assumptions
that are typically hard to test, it is common to instead use cross-validation to
tune \(\alpha\) over a grid of values. In practice, this means that we
repeatedly have to fit the full regularization path, which is the sequence of
solutions to the problem in \autoref{eq:slope} as \(\alpha\) is varied from
\(\alpha_\text{max}\) (the value at which the first cluster enters the model) to
a small value of \(\alpha\) at which the model is almost saturated. Our packages
are optimized to efficiently compute the full regularization path, and make use
of screening rules~(\autoref{sec:screening-rules}) to speed up the process of
doing so.

We use the same criteria as \citet{friedman2010} for stopping the path early,
except that we stop if the number of \emph{clusters}, excluding the
zero-cluster, exceeds \(n + 1\) (by default), since the support of SLOPE is
limited at \(n\) clusters\footnote{As opposed to the lasso, which at most allows
\(n\) non-zero \emph{coefficients}.}. This can potentially exceed the number of
non-zero coefficients, although in practice this is rare since clusters do not
form easily at low levels of regularization.

\subsection{Screening rules}
\label{sec:screening-rules}

Sparse models like lasso and SLOPE are well-known to benefit from
\emph{screening rules}, which are used to reduce the dimension of \(\beta\) for
a fixed \(\lambda\) in the optimization problem and thereby speed up
optimization. The intuition for this is that it is possible to estimate the
gradient \(\nabla F(\beta)\) for a given SLOPE problem and, via the
subdifferential, estimate the support of the solution: the identity of the
nonzero coefficients. Screening rules are either \emph{heuristic} or
\emph{safe}. In the latter case the rule guarantees that excluded predictors
correspond to zero coefficients in the final model. Heuristic rules, on the
other hand, do not guarantee this and therefore need to be complemented with a
pass over all coefficients at the end to ensure that the optimality conditions
are satisfied. But since heuristic rules are typically less conservative, the
computational gains often outweigh this extra cost in practice.

In the SLOPE package, we use the strong screening rule for
SLOPE~\citep{larsson2020a}, which is an extension of the working set strategy
for the strong screening rule for the lasso~\citep{tibshirani2012}.

\section{Implementation details}
\label{sec:implementation-details}

In this section, we provide an overview of the implementation details of the
SLOPE packages, including the data structures used to represent clusters, the
thresholding operator, and parallelization strategies.

\subsection{Software architecture}

We have implemented a collection of packages for solving SLOPE, currently with
support for \proglang{R}, \proglang{Python}, and \proglang{Julia}. The backbone
of all of these packages is based on a \proglang{C++} library that implements
all of the numerical algorithms for SLOPE, including preprocessing,
cross-validation, and path fitting. The packages for the high-level languages
all serve as thin wrappers to the \proglang{C++} library, with some additional
functionality for handling data and plotting the results. This means that new
features and bug fixes propagate quickly and easily to all these wrappers and
enable users to promptly take advantage of the latest developments. The entire
suite of packages is open source and available in version-controlled online
repositories~(\autoref{tab:slope-packages}).

\begin{table}[tp]
  \centering
  \begin{tabular}{llll}
    \toprule
    Language          & Package        & Repository                         & Documentation                     \\
    \midrule
    \proglang{R}      & \pkg{SLOPE}    & \myurl{github.com/jolars/SLOPE}    & \myurl{jolars.github.io/SLOPE}    \\
    \proglang{Python} & \pkg{sortedl1} & \myurl{github.com/jolars/sortedl1} & \myurl{jolars.github.io/sortedl1} \\
    \proglang{Julia}  & \pkg{SLOPE.jl} & \myurl{github.com/jolars/SLOPE.jl} & \myurl{jolars.github.io/SLOPE.jl} \\
    \proglang{C++}    & \pkg{slope}    & \myurl{github.com/jolars/libslope} & \myurl{jolars.github.io/libslope} \\
    \bottomrule
  \end{tabular}
  \caption{
    A summary of the suite of packages that we have developed for solving SLOPE,
    along with links to the source code repositories and documentation.
  }
  \label{tab:slope-packages}
\end{table}

This is made possible via several pieces of software that enable us to link the
API from our \proglang{C++} library to the high-level languages. This includes
\pkg{Rcpp}~\citep{eddelbuettel2011} and \pkg{RcppEigen}~\citep{bates2013} for
\proglang{R}, \pkg{pybind11}~\citep{jakob2025}, and
\pkg{CxxWrap}~\citep{janssens2020} for \proglang{Julia}.

\subsection{Core algorithmic components}

Handling the cluster structure of SLOPE is a key part of the algorithm since we
will both be iterating over the clusters as part of the coordinate descent
updates as well as updating the clusters after each update. In our
implementation, we represent the clusters as a collection of three vectors:

\begin{description}
  \item[\code{c}] The coefficients of the clusters
  \item[\code{c\_idx}] Pointers to the coefficients in the cluster
  \item[\code{c\_ptr}] Values of the cluster pointers
\end{description}

In this representation, the indices for the \(k\)th cluster are given by
\code{
c_idx[c_ptr[k] : c_ptr[k+1]]}
and the coefficient is simply \code{c[k]}. This structure is the same basic
setup as in \citet{larsson2023}. Unlike their implementation, however, we have
made improvements to the handling of updating the clusters (merging, reordering,
removal), which can now be performed with negligible overhead and with minimal
copying.

The SLOPE thresholding operator~(\autoref{fig:slope-thresholding}) is the
analogue to the soft-thresholding operator for the lasso. But unlike the latter,
which is trivial to compute, the SLOPE thresholding operator needs to conduct a
search over the clusters in order to find correct solution. This leads to a
worst-case complexity that depends on the number of clusters, which might seem
to be prohibitive when both \(n\) and \(p\) are large. Fortunately, however, the
situation is much less dire in practice, since the order of the clusters
typically stabilizes early during optimization. In benchmarks, we have found
that it makes no difference whether we use a linear search or a binary search
for real data sets, and we have therefore opted for the simpler linear search.
Note also that the partial \(\lambda\) sums do not need to be computed directly.
Instead, we simply compute the cumulative sum of the \(\lambda\) array once and
use this to retrieve the partial sums as needed.

\subsection{Data handling and optimization}

Our package is based on the \pkg{Eigen} \proglang{C++} library and provides
support for both dense and sparse design matrices. The latter can be constructed
through the \pkg{Matrix}~(\proglang{R}), \pkg{scipy}~(\proglang{Python}), and
\pkg{SparseArrays}~(\proglang{Julia}) packages in and are passed to the
\proglang{C++} API without copying and with negligible overhead. Coefficients
are returned in a sparse format, which allows for efficient storage and
retrieval of the coefficients.

For dense matrices, the packages implement memory-efficient views of the input
matrices to avoid copying the data. This means, for instance, that no copies
need to be made when separating data sets into training and test data. Due to
this memory-efficient implementation, we provide an option to altogether avoid
copying the data during cross-validation. This allows us to parallelize the
cross-validation procedure over arbitrarily many folds and repetitions without
needing to ever copy the data, which makes for a much more memory-friendly
implementation (at the cost of worse runtime performance).

The SLOPE packages also support out-of-memory storage for the design matrix,
which means that users can fit SLOPE models on data sets that are larger than
the available memory. Storage in RAM is therefore limited to the order of
\(O(n) + O(p)\), which means that the packages can be used on huge data sets.
This is supported via the generic \code{Eigen::Map} class, which allows
arbitrary data to be mapped into \pkg{Eigen} data structures. At the time of
writing, this is only supported for \proglang{R}, which is possible via the
\pkg{bigmemory} package~\citep{kane2013}, and currently only for dense designs.

As shown by \citet{larsson2025}, predictor normalization (centering and scaling
the design matrix) may have large consequences for the solutions. Our packages
provide multiple different options for centering and scaling, independently of
one another. We also provide the possibility to manually supply centering and
scaling vectors. Optionally, normalization is also realized just-in-time (JIT),
which means that the design matrix does not need to be normalized in place.
Normalization is performed as predictors of the design matrix are accessed
during optimization, which allows us to support centering even of sparse design
matrices. This also allows us to completely avoid copying the design matrix.

The software is parallelized using \pkg{OpenMP}, which is supported on all major
platforms. Our functions employ several heuristics based on problem size to
determine whether to spawn multiple threads, except for the case of
cross-validation, which is always parallelized.

\section{Examples}
\label{sec:examples}

The packages are available through the respective package managers for each
language and can be installed using the following commands:
\begin{description}[labelwidth=8ex]
  \item[\proglang{R}] \code{R> install.packages("SLOPE")}
  \item[\proglang{Python}] \code{$ pip install sortedl1}
  \item[\proglang{Julia}] \code{julia> using Pkg; Pkg.add("SLOPE")}
\end{description}

Installing the \proglang{C++} library is slightly more involved, and requires
\pkg{CMake}~\citep{kitware2025} together with a working \proglang{C++}
toolchain, including the \pkg{Eigen} library~\citep{guennebaud2010a} and
\pkg{OpenMP}~\citep{dagum1998} (optionally, to enable parallelization).

Assuming we have loaded a data set consisting of a design matrix \code{x} and
response vector \code{y}, we can fit the full regularization path for the SLOPE
model using the following commands in the different languages:

\begin{minipage}[t]{0.24\textwidth}%
  \textbf{\proglang{R}}
  \begin{Code}
R> library(SLOPE)
R> fit <- SLOPE(x, y)
  \end{Code}
\end{minipage}
\hfill
\begin{minipage}[t]{0.33\textwidth}%
  \textbf{\proglang{Python}}
  \begin{Code}
>>> import sortedl1
>>> model = sortedl1.Slope()
>>> fit = model.path(x, y)
  \end{Code}
\end{minipage}
\hfill
\begin{minipage}[t]{0.32\textwidth}%
  \textbf{\proglang{Julia}}
  \begin{Code}
julia> using SLOPE
julia> fit = slope(x, y)
  \end{Code}
\end{minipage}

\medskip

You can also use the C++ library directly, in which case the above would
translate into the following:
\begin{Code}
#include <slope/slope.h>
slope::Slope model;
auto path_result = model.path(x, y);
\end{Code}

In the sequel, we will focus our examples on the \proglang{R} package, but
please see \autoref{sec:examples-julia-python} for equivalent examples in
\proglang{Python}, \proglang{Julia}, and \proglang{C++}.

\subsection{First steps}
\label{sec:first-steps}

We start with a simple example of fitting a full SLOPE path to the diabetes data
set~\citep{efron2004}, including plotting it.\footnote{Note that the actual code
for the plotting examples here is slightly more complex to allow for better
control over the aesthetics.}

\begin{Code}
R> library(SLOPE)
R> data("diabetes", package = "lars")
R> x <- scale(diabetes$x)
R> y <- diabetes$y
R> fit_slope <- SLOPE(x, y, q = 0.4)
R> plot(fit_slope)
\end{Code}

The \code{q} parameter is a parameter of the sequence of \(\lambda\) values,
which by default~(\code{lambda = "bh"}) uses the Benjamini--Hochberg (BH)
sequence~\citep{bogdan2015}. If the design matrix is orthogonal, then the
\code{q} parameter sets a desired false discovery rate (FDR) in terms of the
identification of true signals (nonzero coefficients).

Other types of sequences are also supported, including \code{lambda = "lasso"}
for the lasso, \code{lambda = "oscar"} for the OSCAR
sequence~\citep{bondell2008}, and \code{lambda = "gaussian"} for the
Gaussian-type sequence~\citep{bogdan2015}, which is a modification of the BH
sequence that has been empirically shown to provide similar FDR control in
non-orthogonal, low-dimensional settings.

To show how the choice of the \(\lambda\) sequence affects the results, we will
refit the diabetes data using the lasso sequence.

\begin{Code}
R> fit_lasso <- SLOPE(x, y, lambda = "lasso")
R> plot(fit_lasso)
\end{Code}

\begin{figure}[tp]
  \centering
  {\includegraphics[width=\natwidth]{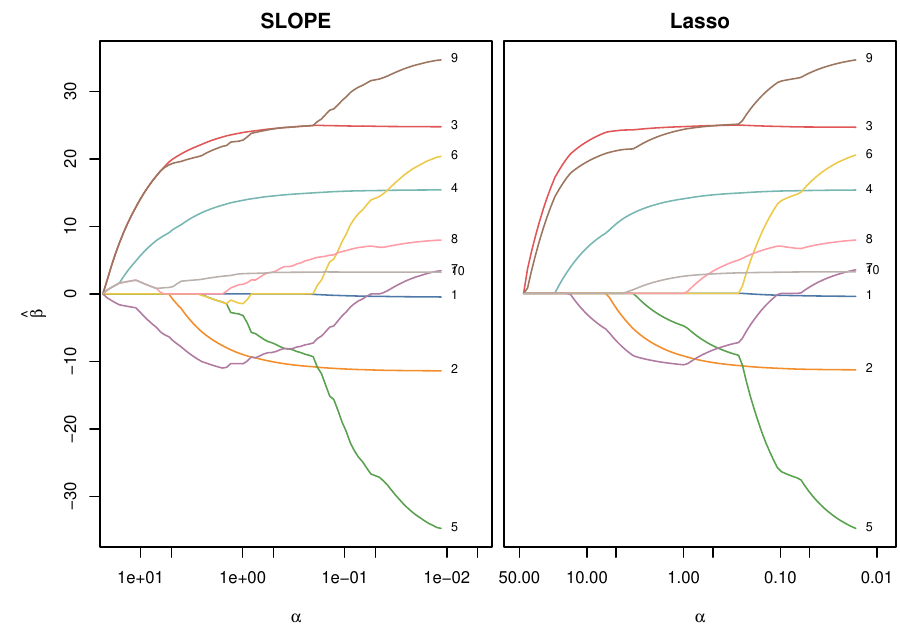}}
  \caption{
    SLOPE and lasso paths on the diabetes data set. Note that the \(x\) axis is
    reversed. Numbers indicate the indices of the predictors.
  }
  \label{fig:diabetes}
\end{figure}

The resulting paths are plotted in \autoref{fig:diabetes}. Observe that the
paths are similar but that SLOPE has clustered some coefficients along parts of
the path. Predictors 3 and 9, for instance, enter the path together and remain
clustered at the beginning, then split apart and again cluster together briefly,
before diverging again. We also see that predictors 4, 10, and 8 enter the model
together, only to then, one by one, split apart later on. Unlike the lasso path,
there are many more \emph{kinks} in the SLOPE path, since they occur not only
when new predictors become active, but also when they merge into or split from
clusters.

\subsection{Relaxed SLOPE}

To attain sparsity, SLOPE shrinks coefficients towards zero. This introduces
bias, which, although it helps to combat overfitting, may also lead to worse
predictions in some cases. To mitigate this, it is possible to \emph{relax} the
SLOPE solutions by fitting an ordinary least-squares model to the collapsed
cluster structure from running SLOPE~\citep{skalski2022}\footnote{Note that this
paper refers to the relaxed SLOPE as the ``debiased'' SLOPE\@. Here, we prefer
the term ``relaxed''' to stay closer to the nomenclature from the lasso
literature and avoid confusion with the debiased lasso~\citep{geer2014}, which
represents a different approach.}. The level of relaxation is parameterized by
\(\gamma\), which controls the mix between the original SLOPE solution
(\(\gamma = 1\)) and the fully relaxed solution (\(\gamma = 0\)), so that the
end result is given by
\[
  \hat{\beta} = \gamma \hat{\beta}_\text{SLOPE}
                + (1 - \gamma ) \hat{\beta}_\text{relaxed} .
\]

Our packages support regularization through the \code{gamma} argument. Here, we
fit fully relaxed~(\(\gamma = 0\)) and semi-relaxed~(\(\gamma = 0.5\)) models,
along with standard SLOPE~(\(\gamma = 1\)) for comparison.

\begin{Code}
R> fit_relaxed <- SLOPE(x, y, q = 0.1, gamma = 0)
R> fit_semirelaxed <- SLOPE(x, y, q = 0.1, gamma = 0.5)
R> fit_standard <- SLOPE(x, y, q = 0.1, gamma = 1)
\end{Code}

\begin{figure}[tp]
  \centering
  \includegraphics[width=\natwidth]{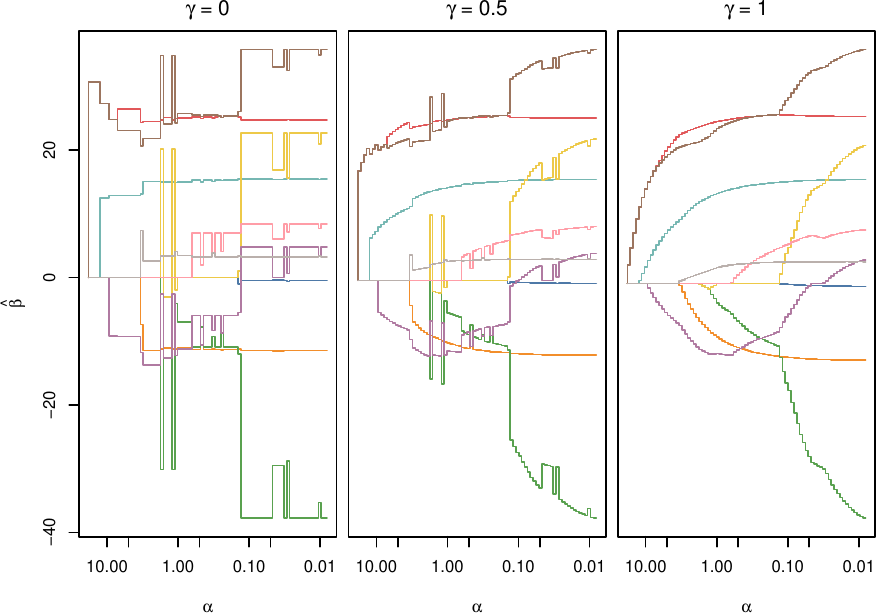}
  \caption{
    SLOPE with various levels of relaxation \(\gamma\), with \(\gamma = 0\)
    (fully relaxed), \(\gamma = 0.5\) (semi-relaxed), and \(\gamma = 1\)
    (standard SLOPE).
  }
  \label{fig:relaxed-slope}
\end{figure}

We plot the resulting paths in \autoref{fig:relaxed-slope}. Observe the
jaggedness of the relaxed paths, which change dramatically as clusters form,
merge, and split.

\subsection{Cross-validation}

Our packages support hyper-parameter tuning via iterated \(k\)-fold
cross-validation (CV), with parameterization over \(\alpha\), \(\lambda\) type
(BH, Gaussian type, etc.), \(\gamma\) (SLOPE relaxation parameter). Here, we
demonstrate the CV functionality by cross-validating across two values of the
\(q\) parameter and printing the results, which displays the optimal values for
all the cross-validated parameters.

\begin{CodeChunk}
  \begin{CodeInput}
R> set.seed(48)
R> fit_cv <- cvSLOPE(x, y, q = c(0.1, 0.2))
R> fit_cv
\end{CodeInput}
  \begin{CodeOutput}
Call:
cvSLOPE(x = x, y = y, q = c(0.1, 0.2))

Optimum values:
    q gamma     alpha measure     mean      se       lo       hi
1 0.2     0 0.3379402     mse 3012.926 234.652 2482.106 3543.745
\end{CodeOutput}
\end{CodeChunk}

\begin{figure}[tp]
  \centering
  \includegraphics[width=\natwidth]{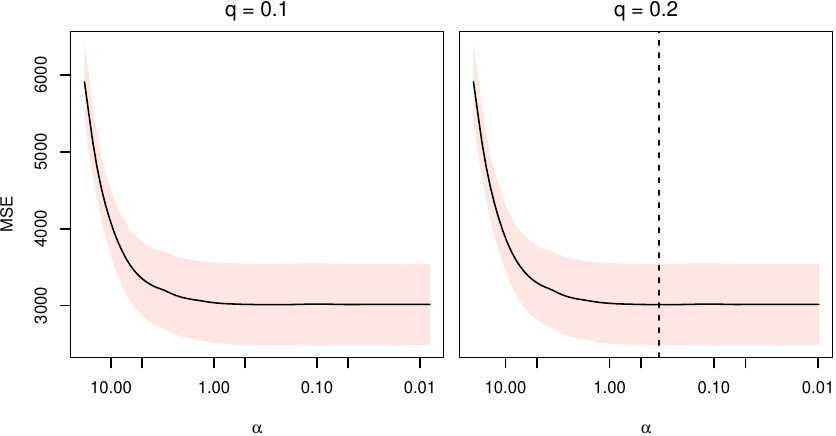}
  \caption{
    Mean-squared error (MSE) from cross-validation of \(q\) and \(\alpha\) for
    SLOPE fit to the diabetes data set. The dashed line marks the optimal value
    of \(\alpha\) in the panel corresponding to the optimal value of \(q\).
  }
  \label{fig:cv}
\end{figure}

It is also easy to plot the cross-validation results~(\autoref{fig:cv}), which
show the cross-validation error with 95\% confidence intervals and a dashed line
indicating the optimal value of \(\alpha\) for the best value of \(q\).

\begin{Code}
R> plot(fit_cv)
\end{Code}

\section{Benchmarks}

In this section, we present benchmarks of the numerical performance of our
implementation for solving SLOPE\@. In \autoref{sec:single-solution-benchmark},
we examine the performance of SLOPE when fitting for a single value of
\(\alpha\), while in \autoref{sec:path-benchmark}, we benchmark the performance
of fitting the full regularization path.

Our benchmarks are organized and run using \pkg{benchopt}~\citep{moreau2022a}.
They are available as public git repositories at
\myurl{github.com/benchopt/benchmark\_slope} and
\myurl{github.com/jolars/benchmark\_slope\_path} for the single solution and
path benchmarks, respectively. Both feature the Python version of our
implementation, \pkg{sortedl1}\footnote{The \proglang{R} package \pkg{SLOPE} is
also included, but we have omitted it from the benchmark here since it is
essentially equivalent to the \proglang{Python} package \pkg{sortedl1}.}. The
single solution benchmark includes four variations on proximal gradient descent
(PGD), using Anderson acceleration~\citep{anderson1965, zhang2020},
Barzilai--Borwein step sizes, safe screening rules, and the fast iterative
shrinking and thresholding algorithm (FISTA)~\citep{beck2009}. The latter of
these is also included through the \pkg{skglm} package~\citep{bertrand2022}. We
also include the alternating direction method of multipliers
(ADMM)~\citep{boyd2010}, a semi-smooth Newton-based method~\citep{luo2019}, and
an approximate homotopy method, \pkg{SolutionPath}~\citep{dupuis2024}. See
\autoref{sec:solver-details} for more details on the solvers and their
implementations. The path benchmark compares \pkg{sortedl1} with a subset of
these solvers, namely \pkg{FISTA}, \pkg{SolutionPath}, and \pkg{ADMM}. All of
the benchmarks were run on a system with an AMD Ryzen 9 7900 CPU (12 cores and
24 threads, with a maximum boost clock of 5.5 GHz), 64 GB of RAM, and NixOS
26.11 (Linux).

We have run the benchmarks for both simulated~(\autoref{tab:simulated-data}) as
well as real data~(\autoref{tab:real-data}). For all real data, we standardize
the predictors to have mean zero and unit variance if \(X\) is dense and scale
with the maximum absolute value of each predictor otherwise. For all data we use
the Benjamini--Hochberg sequence for \(\lambda\) with \(q = 0.2\). Since some of
the solvers cannot handle intercepts, we have omitted the intercepts from the
models. Also note that not all solvers support sparse design matrices, which is
why they are not included everywhere.

We evaluate convergence using a solver-independent relative duality gap and stop
sampling each convergence curve once this measure reaches \(10^{-7}\). For
solution paths, we apply the same target to the maximum relative duality gap
along the path. Values below this threshold are not distinguished in the
figures. The solvers retain their native internal stopping rules, so the common
duality-gap measure, rather than an internal tolerance parameter, provides the
basis for comparison. We use Benchopt's 30-second timeout, which is checked
between successive evaluations of a solver, and restrict the figures to
evaluations taking no more than 30 seconds.

\begin{table}[tp]
  \centering
  \begin{tabular}{
      l
      S[table-format=6.0]
      S[table-format=7.0]
      S[table-format=1.5,round-mode=figures,round-precision=2]
      p{5cm}
    }
    \toprule
    Dataset                     & {\(n\)} & {\(p\)} & {\(X\) density} & {References}                        \\
    \midrule
    \dataset{BRCA1}             & 536     & 17322   & 1               & \citet{nationalcancerinstitute2022} \\
    \dataset{Koussounadis2014}  & 101     & 34694   & 1               & \citet{koussounadis2014}            \\
    \dataset{Real-Sim}          & 72309   & 20958   & 0.00245         & \citet{mccallum2010}                \\
    \dataset{RCV1}              & 20242   & 44504   & 0.00166         & \citet{lewis2004}                   \\
    \dataset{Rhee2006}          & 842     & 360     & 0.02469         & \citet{rhee2006}                    \\
    \dataset{YearPredictionMSD} & 463715  & 90      & 1               & \citet{bertin-mahieux2011}          \\
    \bottomrule
  \end{tabular}
  \caption{
    List of real datasets used in our experiments, along with some of their
    properties, including the number of samples \(n\) and predictors \(p\).
    \dataset{BRCA1}, \dataset{Koussounadis2014}, and \dataset{Rhee2006} were
    obtained from \citet{breheny2022}, and the rest from \citet{chang2016}.
  }
  \label{tab:real-data}
\end{table}

\begin{table}[tp]
  \centering
  \begin{tabular}{
      l
      S[table-format=6.0]
      S[table-format=6.0]
      S[table-format=2.0]
      S[table-format=1.3]
      S[table-format=0.1]
    }
    \toprule
    {Scenario}       & {\(n\)} & {\(p\)} & {\(s\)} & {\(X\) density} & {\(\rho\)} \\
    \midrule
    High Dim         & 100     & 20000   & 20      & 1               & 0.6        \\
    High Dim, Sparse & 2000    & 200000  & 20      & 0.001           & 0.6        \\
    Low Dim          & 20000   & 100     & 40      & 1               & 0.6        \\
    \bottomrule
  \end{tabular}
  \caption{
    Scenarios for the simulated data in our benchmarks. \(\rho\) is the
    auto-correlation between adjacent predictors, \(s\) is the number of nonzero
    true coefficients, and \(n\) and \(p\) are the number of samples and
    predictors, respectively.
  }
  \label{tab:simulated-data}
\end{table}

\subsection{Single solution}
\label{sec:single-solution-benchmark}

We have parameterized our single-solution benchmarks by \(\alpha\) as a fraction
of \(\alpha_\text{max}\) (the value at which the first cluster enters the
model), and run the benchmarks for \(\alpha = \alpha_\text{max} / 2\),
\(\alpha_\text{max} / 10\), and \(\alpha_\text{max} / 50\). We present
convergence using the duality gap.

\begin{figure}[tp]
  \centering
  \includegraphics[width=\natwidth]{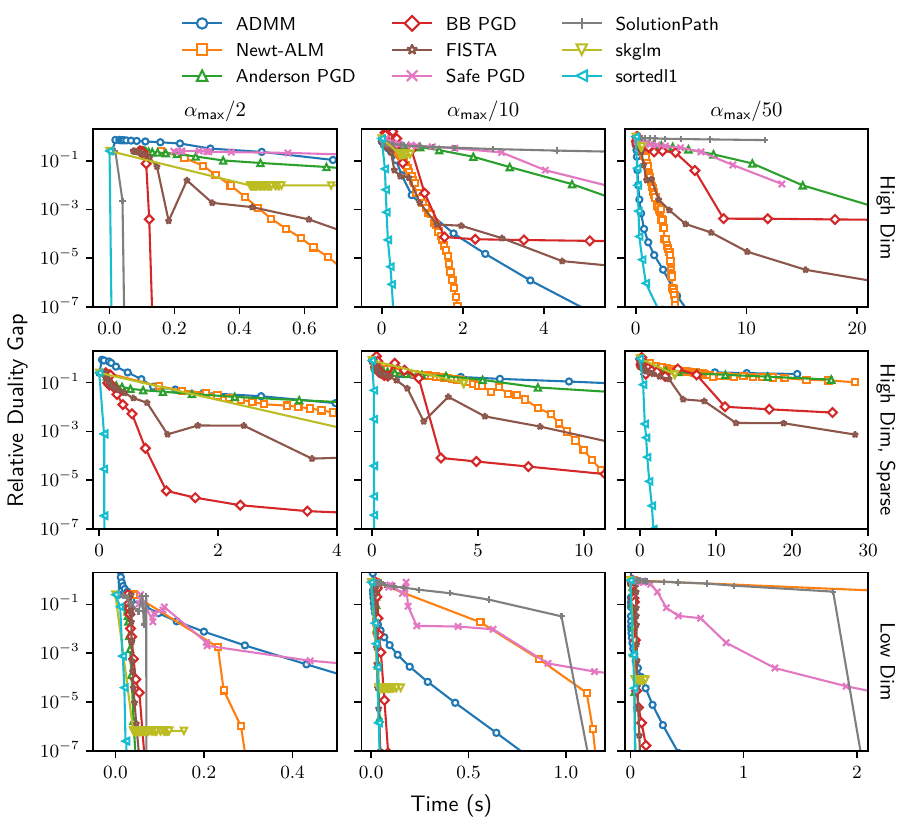}
  \caption{
    Performance of the solvers for the single solution benchmark for three
    different settings of simulated data~(\autoref{tab:simulated-data}). Please
    see the text for information about the data and setup of the experiment.
  }
  \label{fig:simulated-data-single}
\end{figure}

The results for the simulated data are presented in
\autoref{fig:simulated-data-single}. At the displayed relative-duality-gap
threshold of \(10^{-7}\), our algorithm is the first to reach the threshold in
all nine combinations of simulation setting and regularization level. The
difference is especially pronounced for high levels of regularization, where our
method is often many times faster than the next best method. Newt-ALM also
reaches the threshold rapidly in the (\(\alpha_\text{max} / 50\), High Dim)
setting, although \pkg{sortedl1} reaches it first. Our current implementation
shows improved performance compared to \citet{larsson2023}, which also featured
the hybrid method. We believe this is likely due to the addition of screening
rules as well as algorithmic improvements. Newt-ALM also performs better than in
\citet{larsson2023}, following improvements to the implementation used here.

\begin{figure}[tp]
  \centering
  \includegraphics[width=\natwidth]{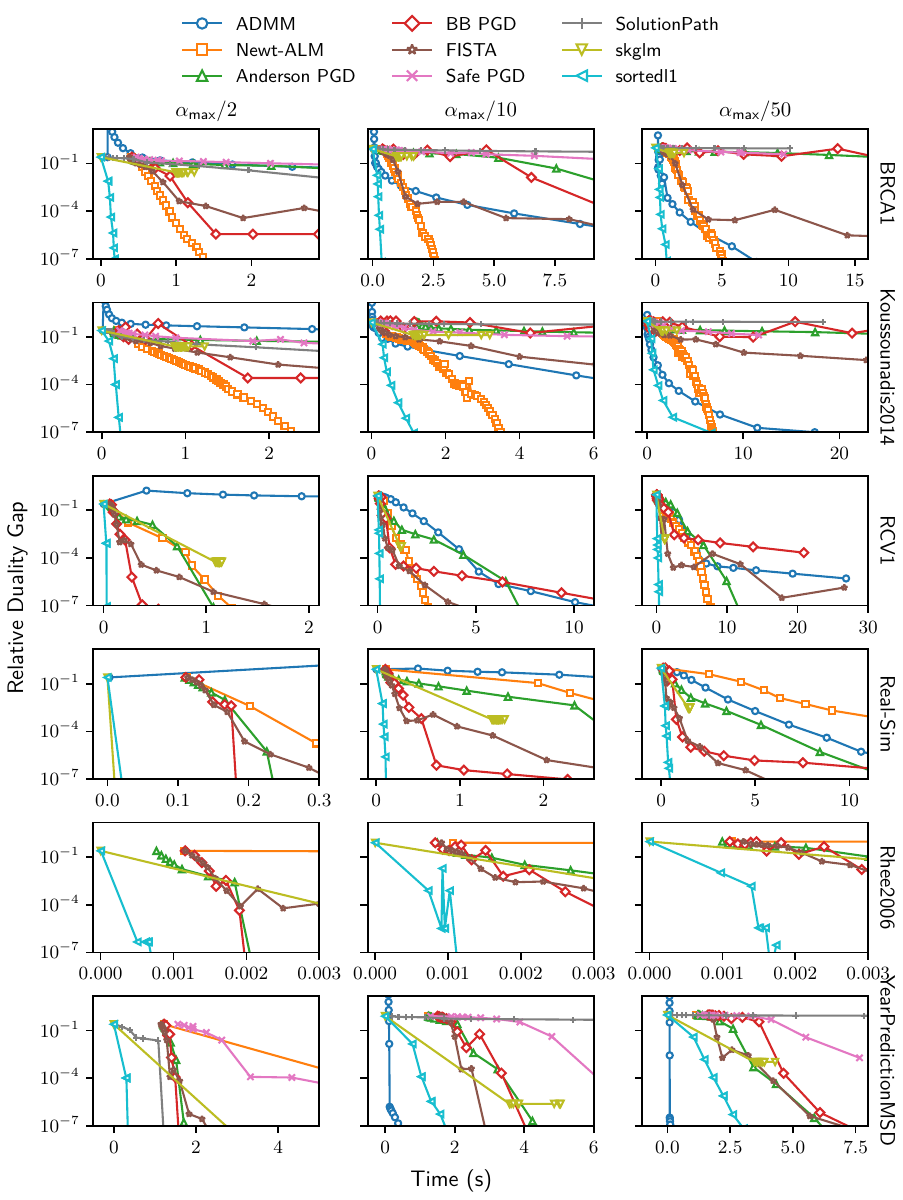}
  \caption{
    Performance of the solvers for the single solution benchmark for six
    different real data sets~(\autoref{tab:real-data}) and three levels of
    regularization. Please see the text for information about the data and setup
    of the experiment.
  }
  \label{fig:real-data-single}
\end{figure}

For real data~\autoref{fig:real-data-single}, we see a mostly similar pattern.
At a relative duality gap of \(10^{-7}\), \pkg{sortedl1} is the first solver to
reach the threshold in 16 of the 18 combinations. ADMM is first for
\dataset{YearPredictionMSD} at \(\alpha_\text{max}/10\) and
\(\alpha_\text{max}/50\). The performance of the other algorithms is generally
more sensitive to the dataset and regularization level.

\subsection{Path}
\label{sec:path-benchmark}

Here, we present benchmarks on a subset of the real data sets from
\autoref{tab:real-data}. We parameterize the benchmark using the length of the
path, using 50, 100, and 200 steps. We present the results as the maximum
relative duality gap along the path.

\begin{figure}[tp]
  \centering
  \includegraphics[width=\natwidth]{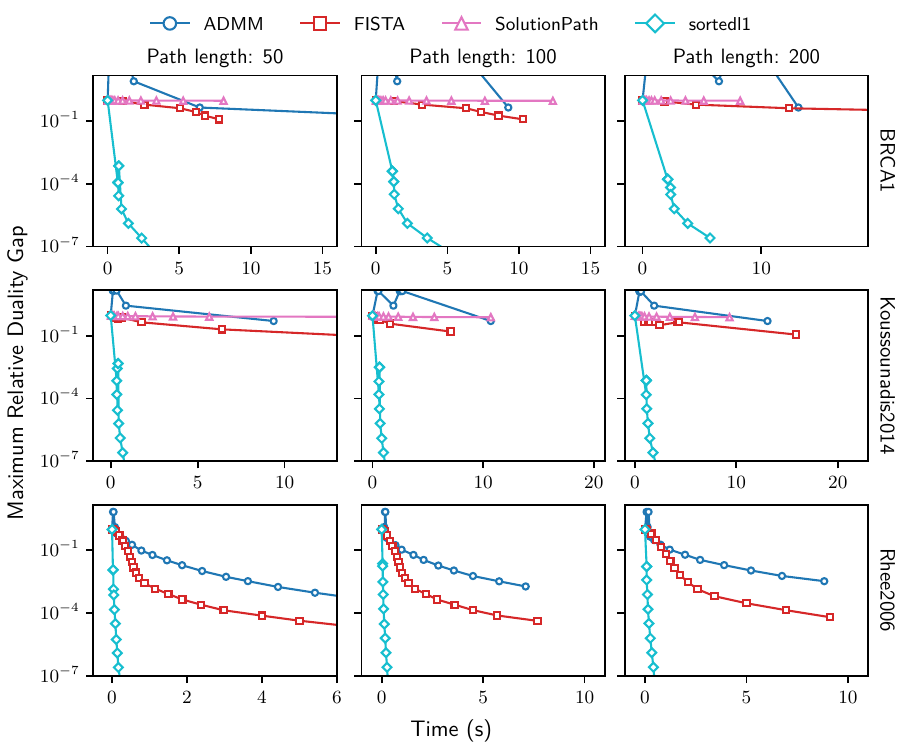}
  \caption{
    Benchmark for fitting the full regularization path on real data sets with
    varying path lengths. The plot shows the maximum relative duality gap across
    all solutions along the regularization path.
  }
  \label{fig:real-data-path}
\end{figure}

The results are presented in \autoref{fig:real-data-path}. Note that the
interpretation of progress towards convergence does not hold quite the same way
as for the single solution benchmarks since we fit a full path. Nevertheless,
\pkg{sortedl1} is the only solver to reduce the maximum relative duality gap to
\(10^{-7}\), reaching it in eight of the nine combinations of dataset and path
length, and doing so by a large time margin. No solver reaches the threshold for
\dataset{BRCA1} with 200 path points within the time limit; \pkg{sortedl1} comes
closest, with a maximum relative duality gap of \(2.5 \times 10^{-7}\).

Taken together, our benchmarks show that our method reaches a relative duality
gap of \(10^{-7}\) first in 25 of the 27 single-solution settings and in eight
of the nine path settings. ADMM is faster in the other two single-solution
settings, while no solver reaches the threshold in the remaining path setting.
In contrast to \citet{dupuis2024}, who showed mixed results when comparing their
approximate homotopy method to the hybrid algorithm from \citet{larsson2023}, we
find a clearer advantage for the current hybrid implementation. Possible
explanations include the more stringent stopping criterion used by
\citet{dupuis2024} (a duality gap of \(10^{-10}\)), the generally larger
problems in our benchmarks, and enhancements over the original hybrid
implementation from \citet{larsson2023}.

\section{Application to real-world data}

To illustrate the practical utility of SLOPE, we applied it to a metabolomics
dataset from \citet{Godlewski2023}, comprising plasma measurements from glioma
patients and healthy controls. The dataset contains 165 samples (94 glioma cases
and 71 controls), each with 138 metabolite features.

We first create the design matrix \texttt{x} and the response vector \texttt{y}:

\begin{Code}
R> x <- glioma$x
R> y <- glioma$y
\end{Code}

We then fit a SLOPE model with cross-validation to select the regularization
parameter. The function \texttt{cvSLOPE()} performs $K$-fold cross-validation
and optimizes over~$\alpha$:

\begin{Code}
R> slope_cv <- cvSLOPE(x, y, q = 0.1, family = "binomial", measure = "auc")
R> alpha_cv <- slope_cv$optima$alpha
R> slope_model <- SLOPE(x, y, q = 0.1, family = "binomial", alpha = alpha_cv)
\end{Code}

Predicted probabilities for new data can be obtained with the \texttt{predict()}
method:

\begin{Code}
R> pred_prob <- predict(slope_model, x, type = "response")
\end{Code}

For illustration, we also split the data into training and test sets. This
allows us to evaluate out-of-sample classification performance when
distinguishing glioma patients from healthy controls:

\begin{Code}
R> set.seed(222)
R> train_index <- caret::createDataPartition(y, p = 0.7, list = FALSE)
R> x_train <- x[train_index, ]
R> y_train <- y[train_index]
R> x_test  <- x[-train_index, ]
R> y_test  <- y[-train_index]
\end{Code}

In this example, SLOPE achieved an AUC of 0.913 on the test set, while the lasso
obtained 0.929. Although the lasso performed slightly better in terms of AUC,
the fitted SLOPE model exhibits the structure induced by the sorted $\ell_1$
penalty: it selected 11 metabolites and organized them into five coefficient
clusters. The lasso selected seven metabolites without grouping them. One of the
unique features of the R implementation is the ability to visualize the
clustering of coefficients with the \texttt{plotClusters()}
function\footnote{This functionality is specific to the R package.}. The plot
highlights groups of variables with coefficients of the same magnitude (up to
sign):

\begin{Code}
R> fit_pat <- SLOPE(x, y, q = 0.1, pattern = TRUE, family = "binomial")
R> plotClusters(fit_pat, include_zeroes = FALSE)
\end{Code}

\begin{figure}[tp]
  \centering
  \includegraphics[width=\textwidth]{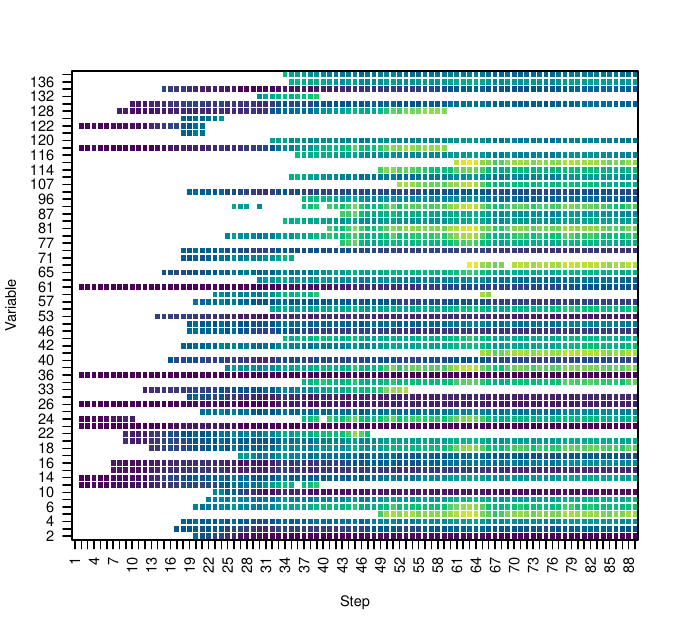}
  \caption{
    Cluster structure of metabolites selected by SLOPE@\. Variables are grouped
    by effect size, with colors indicating distinct clusters.
  }
  \label{fig:real-data-clusters}
\end{figure}

This produces a cluster map of selected metabolites, revealing the grouping
behavior of SLOPE~(\autoref{fig:real-data-clusters}). At the value of~$\alpha$
selected by cross-validation, one cluster comprises \textit{lysine},
\textit{phenylalanine}, and PC aa C34:1, while another comprises
\textit{alanine} and \textit{kynurenine}. The latter is a known glioma
biomarker~\citep{du2020both}, and both \textit{phenylalanine} and
\textit{lysine} have been associated with glioma-related metabolic
changes~\citep{koslinski2023comparative, srivastava2025amino}. The largest
cluster contains \textit{asparagine}, \textit{methionine}, PC ae C42:5, and SM
C18:0. These groupings illustrate how SLOPE can select variables jointly and
assign them coefficients with the same magnitude (up to sign).

Overall, this example demonstrates how the package can be applied in practice.
Beyond predictive performance, its functionality, such as cross-validation,
prediction, and cluster visualization, makes it a useful tool for
high-dimensional biomedical data analysis, where interpretability and feature
selection are essential.

\section{Discussion}
\label{sec:discussion}

SLOPE is an appealing model for high-dimensional regression problems that is
able to recover sparsity and ordering patterns in the solution, which sets it
apart from similar models such as the lasso and elastic net. Unlike these models
however, SLOPE represents a more challenging optimization problem due to the
penalty term's inseparability.

Despite the complexity of the problem, we have here presented a collection of
packages (in \proglang{R}, \proglang{Python}, \proglang{Julia}, and
\proglang{C++}) that offer efficient and feature-rich implementations of
SLOPE\@. We hope that this endeavor will make SLOPE accessible to a wider
audience. Other implementations include \pkg{skglm} and the specialized
approximate path method \pkg{SolutionPath}; our suite instead provides a common
implementation and feature set across four programming languages.

Our benchmarks show that the software generally performs strongly compared with
the other algorithms and implementations considered here, both for fitting the
full regularization path and for a single solution on real and simulated data.

Our results support the effectiveness of the hybrid algorithm approach, with
performance improvements over \citet{larsson2023} following the addition of
screening rules and other algorithmic enhancements. The method reaches the
reported accuracy threshold first in most of the settings considered here,
including eight of the nine path-fitting settings, while another solver is
faster in two single-solution settings. No solver reaches the threshold in the
remaining path-fitting setting.

Although our packages are full-fledged implementations of SLOPE, there are still
some features that are missing and feature coverage is generally not on par with
that of, for instance, \pkg{glmnet} for elastic net. For instance, we do not yet
support observation weights or the full suite of loss functions from the family
of generalized linear models. On the other hand, we \emph{do} support
out-of-memory storage and much more granular normalization options, which are
not supported in \pkg{glmnet}. We also do not yet support the group sorted
\(\ell_1\) norm, which would allow for an alternative penalization scheme for
multivariate response problems. In addition, several possible improvements to
the hybrid method could be considered, such as accelerated and parallelized
coordinate steps. We leave these possibilities to future work but want to stress
that our packages are modular and have been designed with extensibility in mind,
which we hope will facilitate the addition of new features in the future.
Because all of the packages rely on the same \proglang{C++} library,
contributions will also propagate directly to the higher-level packages in
\proglang{R}, \proglang{Julia}, and \proglang{Python}.

We hope that our packages will be useful for researchers and practitioners alike
and that the design of our software suite might inspire others to more closely
couple the available features of \proglang{R}, \proglang{Python}, and
\proglang{Julia}, and avoid redundant implementations of the same algorithms in
each language.

\section*{Acknowledgments}

This research was also supported in part by the French National Research
Agency~(ANR) through the BenchArk project~(ANR-24-IAS2-0003) and by the National
Science Centre Poland through the grant no. 2021/43/O/ST6/02805.

\bibliography{main}

\newpage

\begin{appendix}

\section{Examples in Julia, Python, and C++}%

\label{sec:examples-julia-python}

In this section, we provide \proglang{Python}, \proglang{Julia}, and
\proglang{C++} equivalents of the \proglang{R} examples from
\autoref{sec:first-steps}. As in the \proglang{R} case, some of the plotting
code has been simplified for clarity. Please refer to the code repository for
the full code.

\subsection{Julia}

We begin by loading the necessary packages.

\begin{Code}
julia> using SLOPE
julia> using CSV
julia> using ProjectRoot
julia> using DataFrames
julia> using Statistics
julia> using Random
julia> using Plots
  \end{Code}

Next, we load the diabetes data set, which we have made available as a CSV file
in the \texttt{data} directory of the repository. As before, we standardize the
features to have mean zero and unit variance.

\begin{Code}
julia> df = CSV.read(file, DataFrame)
julia> x = Matrix(df[:, 2:end])
julia> y = df[:, 1]

julia> x_means = mean(x, dims = 1)
julia> x_std = std(x, dims = 1)
julia> x = (x .- x_means) ./ x_std
\end{Code}

We then fit SLOPE and lasso models to the data. Note that the \code{lambda}
argument is actually the symbol $\lambda$ in the \proglang{Julia} package, so
this code will not work if you write \code{lambda = "lasso"}.

\begin{Code}
julia> fit_lasso = slope(x, y, lambda = :lasso)
julia> fit_slope = slope(x, y, lambda = :bh)
  \end{Code}

\begin{figure}[htpb]
  \centering
  \includegraphics[width=\natwidth]{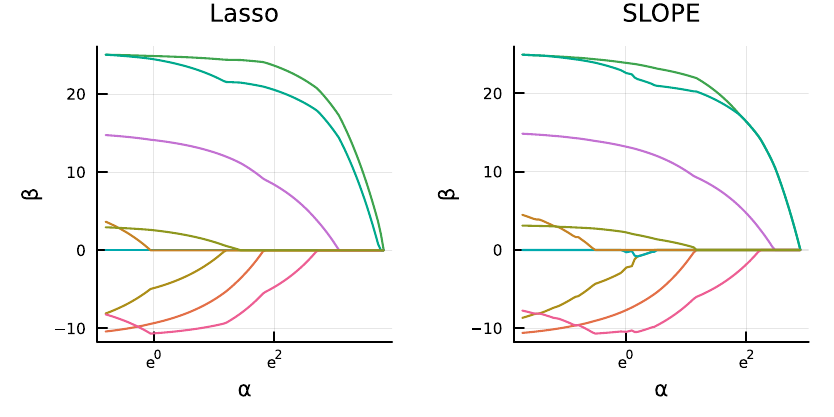}
  \caption{
    SLOPE and lasso paths on the diabetes data set. Note that the \(x\) axis is
    reversed.
  }
  \label{fig:diabetes-julia}
\end{figure}

Just like the \proglang{R} package, the \proglang{Julia} package supports
plotting the solution paths directly. Here, we create a combined plot showing
both the lasso and SLOPE paths side by side~(\autoref{fig:diabetes-julia}).

\begin{Code}
julia> plot(
julia>     plot(fit_lasso)
julia>     plot(fit_slope)
julia>     layout = (1, 2),
julia>     size = (400, 200),
julia>     title = ["Lasso" "SLOPE"],
julia> )
  \end{Code}

To perform cross-validation, we use the \code{slopecv} function, which also
features a dedicated \code{plot} method\footnote{Note that our package only
provides a recipe for the \pkg{Plots.jl} package, and does not depend on it
directly. Users need to load the \pkg{Plots.jl} package (as we do here) to have
access to these plotting methods.}.

\begin{Code}
julia> Random.seed!(48)
julia> fit_cv = slopecv(x, y)
julia> pcv = plot(fit_cv)
  \end{Code}

\begin{figure}[htpb]
  \centering
  \includegraphics[width=\natwidth]{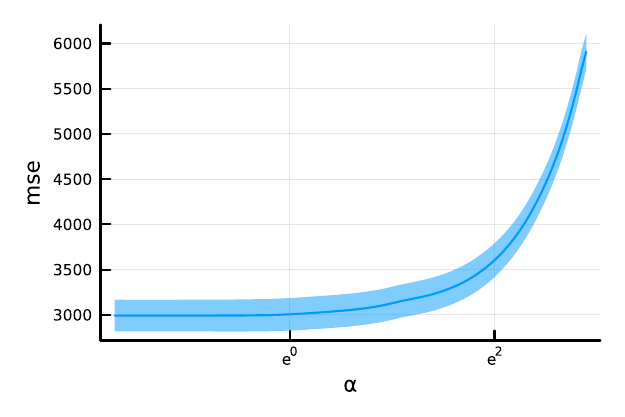}
  \caption{
    Mean-squared error (MSE) from cross-validation over \(\alpha\), with
    \(q = 0.1\), for SLOPE fit to the diabetes data set. The ribbon represents a
    95\% confidence interval.
  }
  \label{fig:cv-julia}
\end{figure}

The result is shown in \autoref{fig:cv-julia}.

\subsection{Python}

We begin by loading the necessary packages.

\begin{Code}
>>> from matplotlib import pyplot as plt
>>> from sklearn.datasets import load_diabetes
>>> from sklearn.preprocessing import StandardScaler
>>> from sortedl1 import Slope
  \end{Code}

Next, we load the diabetes data set from \pkg{scikit-learn} and standardize the
features to have mean zero and unit variance.

\begin{Code}
>>> x, y = load_diabetes(return_X_y=True)
>>> scaler = StandardScaler()
>>> x = scaler.fit_transform(x)
  \end{Code}

We then create SLOPE models with lasso and BH penalty sequences.

\begin{Code}
>>> model_lasso = Slope(lambda_type="lasso")
>>> model_slope = Slope(lambda_type="bh", q=0.4)
  \end{Code}

We fit the models to the data using the \code{path} method.

\begin{Code}
>>> fit_lasso = model_lasso.path(x, y)
>>> fit_slope = model_slope.path(x, y)
  \end{Code}

\begin{figure}[htpb]
  \centering
  \includegraphics[width=\natwidth]{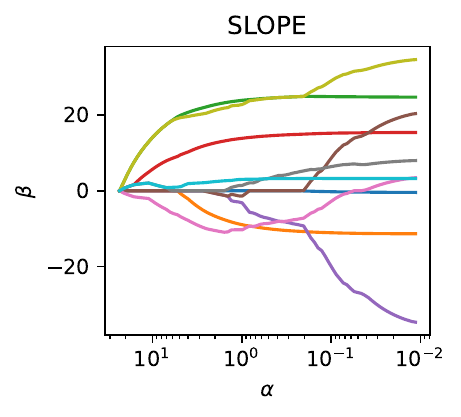}%
  \includegraphics[width=\natwidth]{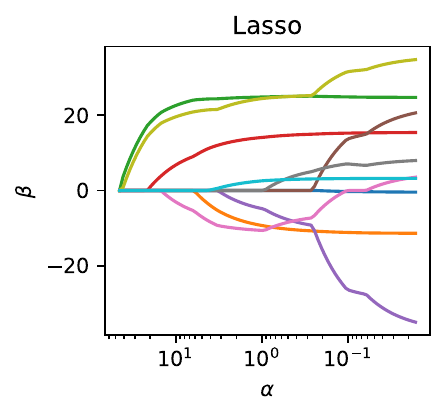}%
  \caption{
    SLOPE and lasso paths on the diabetes data set. Note that the \(x\) axis is
    reversed.
  }
  \label{fig:diabetes-python}
\end{figure}

Just like the \proglang{R} and \proglang{Julia} packages, the \proglang{Python}
package supports plotting the solution paths. The result is shown in
\autoref{fig:diabetes-python}.

\begin{Code}
>>> fit_lasso.plot()
>>> plt.title("Lasso")
>>> plt.show()
>>>
>>> fit_slope.plot()
>>> plt.title("SLOPE")
>>> plt.show()
  \end{Code}

To perform cross-validation, we use the \code{cv} method.

\begin{Code}
>>> fit_cv = model_slope.cv(x, y, q=[0.1, 0.2])
>>> fit_cv.plot()
  \end{Code}

\subsection{C++}

Finally, we provide an example in \proglang{C++} with similar functionality.
Since our \proglang{C++} library does not have built-in plotting capabilities,
nor serialization functionality, we focus here on fitting the models and
printing the coefficients to the console.

\begin{Code}
#include <Eigen/Dense>
#include <iostream>
#include <slope.h>

int main() {
  // Sample data
  Eigen::MatrixXd x(10, 3);
  Eigen::VectorXd y(10);

  x << 1, 2, 3,
       4, 5, 6,
       7, 8, 9,
       1, 0, 1,
       0, 1, 0,
       2, 1, 3,
       3, 2, 1,
       4, 0, 2,
       0, 3, 1,
       1, 1, 1;

  y << 1, 2, 3, 1, 0, 2, 1, 3, 0, 1;

  // Instantiate the SLOPE model
  slope::Slope model;

  // Lasso
  model.setLambdaType("lasso");
  auto fit_lasso = model.path(x, y);

  // SLOPE
  model.setLambdaType("bh");
  model.setQ(0.2);
  auto fit_slope = model.path(x, y);

  // Get the last coefficients on the path
  Eigen::VectorXd coef_lasso = fit_lasso.getCoefs().back();
  Eigen::VectorXd coef_slope = fit_slope.getCoefs().back();

  std::cout << "Lasso Coefficients:" << std::endl << coef_lasso << std::endl;
  std::cout << "SLOPE Coefficients:" << std::endl << coef_slope << std::endl;

  // Cross-validation
  auto cv_res = slope::crossValidate(model, x, y);
  auto best_params = cv_res.best_params;

  std::cout << "\nBest parameters from Cross-validation:" << std::endl;
  for (const auto &param : best_params) {
    printf("
  }

  return 0;
}
  \end{Code}

If the code above is saved in a file named
\texttt{code/example.cpp}\footnote{This is included in the supplementary
material as well.}, it can be compiled with the following command, provided that
GCC is installed along with the \pkg{slope} and \pkg{Eigen3} libraries.
\begin{Code}
g++ -std=c++17 code/example.cpp -lslope -o example 
  \end{Code}

If the resulting executable is run, it produces the following output:

\begin{CodeChunk}
  \begin{CodeInput}
./example
  \end{CodeInput}
  \begin{CodeOutput}
Lasso Coefficients:
 0.385777
-0.397482
 0.326973
SLOPE Coefficients:
 0.385853
-0.397463
 0.326919

Best parameters from Cross-validation:
alpha : 0.000154744
gamma : 0
q     : 0.1
    \end{CodeOutput}
\end{CodeChunk}

\section{Duality gap and convergence criteria}
\label{sec:convergence-criteria-details}

For clarity, we first use the scalar-response notation from
\autoref{sec:math-details}; the multinomial extension follows below. We absorb
\(\alpha\) into \(\lambda\), introduce the linear predictors \(\eta_i\), and
write the primal problem in constrained form:
\begin{equation}
  \begin{aligned}
    \minimize_{\beta_0 \in \mathbb{R},\,\beta \in \mathbb{R}^p,\,\eta \in \mathbb{R}^n} \quad & \frac{1}{n} \sum_{i=1}^n f(\eta_i, y_i) + J_\lambda(\beta)       \\
    \text{subject to} \quad                                                                   & \eta_i = \beta_0 + x_i^\intercal \beta, \qquad i = 1, \ldots, n.
  \end{aligned}
\end{equation}
Using multipliers \(\delta_i/n\) gives the Lagrangian
\[
  L(\beta_0, \beta, \eta, \delta) = \frac{1}{n} \sum_{i=1}^n \left[ f(\eta_i, y_i) - \delta_i(\eta_i - x_i^\intercal\beta - \beta_0) \right]
                                    + J_\lambda(\beta).
\]
Let \(f_i^*\) denote the Fenchel conjugate of \(f_i(\eta) = f(\eta,y_i)\).
Minimizing over \(\eta\) gives
\[
  \inf_{\eta_i}\{f_i(\eta_i) - \delta_i\eta_i\} = -f_i^*(\delta_i).
\]
The terms involving \(\beta\) give
\[
  \inf_\beta \left\{ J_\lambda(\beta) + \left( \frac{X^\intercal\delta}{n} \right)^\intercal\beta \right\} = -J_\lambda^*\left( -\frac{X^\intercal\delta}{n} \right).
\]
Here, the Fenchel conjugate of the sorted \(\ell_1\) norm is the indicator of
its dual-norm unit ball:
\[
  J_\lambda^*(z) = \begin{cases}
                     0      & \text{if } \lVert z \rVert_{*,\lambda} \leq 1, \\
                     \infty & \text{otherwise},
                   \end{cases}
\]
where the sorted \(\ell_1\) dual norm is~\citep{negrinho2014}
\begin{equation}
  \lVert z \rVert_{*,\lambda} = \max_{j=1,2,\dots,p} \left\{ \frac{\sum_{k=1}^j|z_{(k)}|}{\sum_{k=1}^j\lambda_k} \right\}.
\end{equation}
Finally, minimizing over the unpenalized intercept is finite only when
\(\bm{1}^\intercal\delta = 0\). The dual problem is therefore
\begin{equation}
  D(\delta)
  = \begin{cases}
      -\dfrac{1}{n}\displaystyle\sum_{i=1}^n f_i^*(\delta_i) & \text{if } \left\lVert -\dfrac{X^\intercal\delta}{n} \right\rVert_{*,\lambda} \leq 1,\; \bm{1}^\intercal\delta = 0, \\
      -\infty                                                & \text{otherwise}.
    \end{cases} \label{eq:slope-dual}
\end{equation}
As usual, \(f_i^*(\delta_i) = \infty\) outside its domain, so domain feasibility
is implicit in \autoref{eq:slope-dual}. If the model has no intercept, the
condition \(\bm{1}^\intercal\delta = 0\) is omitted.

For the canonical generalized linear models considered here, let
\(q_i = y_i + \delta_i\). The conjugate domain requires \(q_i\) to belong to the
mean-parameter domain \(\mathcal{M}\). This domain is \(\mathbb{R}\) for the
Gaussian model, \([0,1]\) for the binomial model, \([0,\infty)\) for the Poisson
model, and
\[
  \mathcal{M} = \left\{ q \in \mathbb{R}^{m-1}: q_k \geq 0,\; \sum_{k=1}^{m-1}q_k \leq 1 \right\}
\]
for the non-redundant multinomial model. In the multinomial case, we define the
reference-class mean as \(q_{im} = 1 - \sum_{k=1}^{m-1}q_{ik}\). Using the
convention \(0\log 0 = 0\), the loss-specific terms in \autoref{eq:slope-dual}
are
\[
  -f_i^*(\delta_i)
  = \begin{cases}
      -y_i\delta_i - \frac{1}{2}\delta_i^2         & \text{Gaussian},    \\
      -q_i\log q_i - (1 - q_i)\log(1 - q_i)        & \text{binomial},    \\
      q_i(1 - \log q_i)                            & \text{Poisson},     \\
      -\displaystyle\sum_{k=1}^m q_{ik}\log q_{ik} & \text{multinomial}.
    \end{cases}
\]
For example, the Gaussian dual objective is
\[
  D(\delta) = \frac{\lVert y\rVert_2^2 - \lVert y + \delta\rVert_2^2}{2n}
\]
at a feasible point.

The generalized residual \(r_i = \ilink(\beta_0 + x_i^\intercal\beta) - y_i\) is
a natural dual candidate: at the optimum, the Karush--Kuhn--Tucker conditions
include
\[
  \bm{0} \in \frac{X^\intercal r}{n}
             + \partial J_\lambda(\beta), \qquad \bm{1}^\intercal r
         =   0.
\]
Away from the optimum, however, centering \(r\) is safe only for the Gaussian
model. For the other models, centering can move \(y_i + r_i\) outside
\(\mathcal{M}\). We instead construct a zero-sum point that remains in the
conjugate domain. Let
\[
  c_i = r_i - \bar r, \qquad a_i = \bar y - y_i,
\]
and define
\[
  t_{\max} = \max\left\{ t \in[0,1] : y_i + a_i + t(c_i - a_i) \in \mathcal{M} \text{ for every }i \right\}.
\]
Then
\[
  \widetilde\delta_i = a_i + t_{\max}(c_i - a_i)
\]
satisfies both \(\bm{1}^\intercal\widetilde\delta = 0\) and
\(y_i + \widetilde\delta_i \in \mathcal{M}\). The set defining \(t_{\max}\) is
nonempty because \(t = 0\) gives \(y_i + a_i = \bar y \in \mathcal{M}\). For a
model without an intercept, we instead take \(\widetilde\delta = r\), which is
already in the conjugate domain.

It remains to enforce the sorted \(\ell_1\) dual constraint. Set
\[
  \rho = \max\left\{ 1, \left\lVert -\frac{X^\intercal\widetilde\delta}{n} \right\rVert_{*,\lambda} \right\}, \qquad \delta
       = \frac{\widetilde\delta}{\rho}.
\]
This scaling preserves every other constraint. In particular,
\[
  y_i + \delta_i =   \left( 1 - \frac{1}{\rho} \right)y_i
                     + \frac{1}{\rho}(y_i + \widetilde\delta_i)
                 \in \mathcal{M}
\]
by convexity of \(\mathcal{M}\). For binomial and Poisson models, our
implementation obtains a tighter initial point by conditionally adjusting the
intercept so that the residuals sum to zero. It uses the domain-preserving
construction above as a numerical repair and as the primary construction for the
multinomial model. For multinomial regression, the intercept condition holds
componentwise across the nonreference classes, whereas domain feasibility is
imposed jointly on each row through the simplex \(\mathcal{M}\). The regularizer
constraint is also joint: the matrix \(X^\intercal\widetilde\delta/n\) is
vectorized before evaluating the sorted \(\ell_1\) dual norm.

At this feasible point, the duality gap
\[
  P(\beta_0,\beta) - D(\delta)
\]
is nonnegative and bounds the primal suboptimality. We stop when the relative
duality gap satisfies
\[
  P(\beta_0,\beta) - D(\delta) \leq \varepsilon\lvert P(\beta_0,\beta)\rvert,
\]
where \(\varepsilon > 0\) is the user-defined tolerance level. The availability
of the duality gap would also allow us to employ duality-gap based safe
screening rules~\citep{fercoq2015} and working set strategies derived from
these~\citep{massias2018}, which could furthermore be used to enhance our
strategy with look-ahead screening rules~\citep{larsson2021a}. However, as noted
in \citet{larsson2022d}, the marginal improvement of using duality-based
screening strategies is minor, so we have opted not to implement these in our
packages.

\section{Solver details}
\label{sec:solver-details}

\begin{description}
  \item[sortedl1] The python package of our implementation of the hybrid
        proximal gradient/coordinate descent algorithm described in this work
        and \citet{larsson2023}. We use the randomized version of the coordinate
        descent updates.

  \item[Anderson PGD] The proximal gradient descent (PGD) algorithm with
        Anderson acceleration, which is a method for accelerating the
        convergence of fixed-point iterations~\citep{anderson1965, zhang2020}.

  \item[BB PGD] PGD with Barizilai--Borwein~\citep{barzilai1988} step sizes.

  \item[FISTA] The fast iterative shrinking and thresholding
        algorithm~\citep{beck2009}, which is an accelerated version of iterative
        soft-thresholding algorithm (ISTA)~\citep{wright2009}.

  \item[Safe PGD] FISTA with acceleration using safe screening
        rules~\citep{elvira2023}.

  \item[ADMM] The alternating direction method of
        multipliers~\citep{glowinski1975, boyd2010}, which is a popular
        algorithm for solving convex optimization problems with constraints. In
        our experiments, we have used \(\rho = 100\) as a step size based on the
        results from \citet{larsson2023}. We considered using the heuristic
        \(\rho\) selection method from \citet{boyd2010}, but have avoided doing
        so since it, as shown by \citet{larsson2023}, may lead to erratic
        convergence behavior in practice. We use \fct{lsqr} from
        \pkg{SciPy}~\citep{virtanen2020} to solve the linear system if
        \(\min\{n, p\} > \num{1000}\) and otherwise solve the linear system as
        suggested by \citet{boyd2010}.

  \item[Newt-ALM] A semi-smooth Newton-based method~\citep{luo2019}. Our
        implementation of this method is based on that from \citet{larsson2023},
        but improved with better heuristics for selecting the inner solver,
        improved conjugate gradient solver, and a bug fix for the Woodbury-based
        solver. For the conjugate gradient solver, we use \fct{cg} from
        \pkg{SciPy}~\citep{virtanen2020}.

  \item[skglm] Another implementation of FISTA from the \pkg{skglm}
        package~\citep{bertrand2022}.

  \item[SolutionPath] An approximate homotopy method by \citet{dupuis2024},
        which is similar to the lars algorithm for the lasso~\citep{efron2004}.
\end{description}

\end{appendix}

\end{document}